\documentclass[11pt]{article}
\usepackage{graphicx}
\usepackage{amssymb}
\usepackage{amsmath}
\usepackage{epsfig} 
\usepackage[]{caption}
\usepackage{verbatim}
\usepackage{amsfonts}
\usepackage{amsthm}
\usepackage{enumerate}
\usepackage{float}
\usepackage{morefloats}
\usepackage{authblk}
\usepackage{algorithm}
\usepackage{algpseudocode}

\title{The effect of social groups on the dynamics of bi-directional pedestrian flow: a numerical study}

\author[1]{Francesco Zanlungo\thanks{zanlungo@atr.jp}}
\author[2]{Luca Crociani}
\author[1,3]{Zeynep Y{\"u}cel}
\author[1,4]{Takayuki Kanda}

\affil[1]{ATR International, Kyoto, Japan}
\affil[2]{Bicocca University, Milan, Italy}
\affil[3]{Okayama University, Okayama, Japan}
\affil[4]{Kyoto University, Kyoto, Japan}

\begin{document}  
\maketitle
\section*{Abstract}
{\footnotesize \it
Pedestrian social groups represent a large portion of urban crowds and are expected to have a
considerable effect on their dynamics. In this work we investigate the effect of groups on a bi-directional flow, by using novel computational methods.
Our focus is on self-organisation phenomena, and more specifically on the time needed for the occurrence of pedestrian lanes, their
stability and their effect on the velocity-density relation. Moreover, we are interested in understanding the amount of physical contact in the crowd.
To this end, we use a novel model considering the asymmetrical shape of the human body and describing its rotation during collision avoidance, and combine it 
to previously published mathematical model of group behaviour, based on real world uncontrolled
observations of pedestrians in low to moderate densities. We configure several scenarios by varying the
global density $\rho$ of pedestrians and the ratio $r_g$ describing the percentage of grouped pedestrians in
the simulation.
Our results show that the presence of groups has a significant effect on velocity and lane organisation, and a dramatic one on collision. Specifically, crowds with groups walk with slower velocity
(excluding at very high densities),  need more time to organise in lanes and have less pedestrians in lanes; the number of lanes is also
significantly different, at least for some value of $\rho$. As stated, the most dramatic effect is on collisions, which are increased of an order of magnitude in presence of groups.
We are well aware of the limitations of our approach, in particular concerning (i) the lack of calibration of body rotation in collision avoidance on actual data and (ii) straightforward
application of a low density group model to higher density settings. We nevertheless want to stress that it is not our intention to state that our results reproduce the actual effect
of groups on bi-directional flow. In particular, it seems highly unrealistic that crowds with groups collide extremely more often. Nevertheless we believe that our results show the
great theoretical and practical implication of the consideration of realistic group behaviour in pedestrian models, and suggest that realistic results may hardly be achieved simply by
adding together modular models.
}

\newpage
\section{Introduction}

  Pedestrian physical crowds, i.e. not sharing a social identity \cite{reich}, are nevertheless characterised by the presence of a large number of social groups \cite{schultz2}. As a result, it has to be expected that the presence
  of such groups, i.e. of pedestrians that move together and with peculiar velocity pattern and spatial formation \cite{M,M2}, may strongly influence the dynamics of the crowd \cite{he}.
  Although a few microscopic models of group have been recently introduced, \cite{schultz2,M,mou2,costa,zan3,dyads,vizzarri,koster,koster2,wei,kara,zhang,seyfried,gocrofe,zhao,kruchten,wangside,huang,man,kru2,faria,temple}, a quantitative assessing
  of the effect of groups on crowd dynamics is still far from being achieved. This is also due to the fact that, related to the lack of quantitative empirical data, we are still lacking realistic models of how groups behave {\it at different density ranges} (not to mention
  how they behave under different conditions e.g. commuting vs shopping vs evacuation; \cite{intrinsic,intrinsic2} show that purpose, relation, gender, age and height all have a strong impact on group dynamics).

  The purpose of this work is not to provide a definitive answer to these problems, but to show and investigate their relevance. To this end, we are going to
  \begin{itemize}
  \item use a realistic mathematical model for the behaviour of pedestrian groups in low to moderate density settings,
  \item combine it to a realistic and efficient collision avoidance model
  \item and use the resulting model to investigate the effect of the presence of groups on self-organisation properties of a bi-directional flow (velocity/fundamental diagram, amount of collisions,
    number of lanes and rate of pedestrians moving in lanes; these quantities being analysed both in their transition and equilibrium).
  \end{itemize}
  
  \section{Group behaviour}
In \cite{M},  we introduced a mathematical model, based on a pair-wise interaction potential, describing the dynamics of socially interacting pedestrian groups.
Writing the relative position between two socially interacting pedestrians $i$ and $j$ as $\mathbf{r}_{ij}\equiv \mathbf{r}_i-\mathbf{r}_j=(r_{ij},\theta_{ij})$, where $\theta=0$ gives the direction to the pedestrians' goal,
the discomfort of $i$ due to not being located in the optimal position for social interaction with $j$ is given by the (discomfort) {\it potential}
\footnote{In Eq.~\ref{pot}, we are assuming $-\pi < \theta < \pi$, and using
$\text{sign}(0)=-1$ in order to have a continuous potential. Refer to the original work for details.}

\begin{equation}
\label{pot}
\begin{split}
U^\eta(r_{ij},\theta_{ij})= &R(r_{ij}) + \Theta^\eta(\theta_{ij}),\\
R(r)=C_r &\left(\frac{r}{r_0}+\frac{r_0}{r}\right),\\
\Theta^\eta(\theta)=C_\theta \left((1+\eta) \theta^2\right.+&\left.(1-\eta) (\theta-\text{sign}(\theta) \pi)^2\right),
\end{split}
\end{equation}
where $r_0$ is the most comfortable interaction distance, and $-1 \leq \eta < 0$ is related to the intensity of social interaction.
The acceleration of the pedestrian $i$ due to group dynamics, i.e.~the action of the pedestrian aimed to minimise social interaction discomfort with respect to $j$,
 is given by 
\begin{equation}
\label{acc}
\mathbf{F}^{first}_{ij}=-\mathbf{\nabla}_i U^\eta(\mathbf{r}_{ij}).
\end{equation}
In the potential of Eq.~\ref{pot}, the radial term $R$ assures that the pedestrians will have a distance close to $r_0$, while the angular potential $\Theta^{\eta}$ allows them to keep both their interaction partner and their walking goal in sight (the more negative $\eta$ is, the more pedestrians will try to have 
interaction partners in their vision field).
In a pedestrian group, first neighbours interact through the force of Eq. \ref{acc}, while second neighbours interact through a weaker and simplified potential depending only on the radial potential of Eq.~\ref{acc2} \cite{M3},

\begin{equation}
\label{acc2}
\mathbf{F}^{second}_{ij}=-\frac{1}{2} \mathbf{\nabla}_i R(r_{ij}). 
\end{equation}

In this model, a pedestrian's neighbourhood is decided based on its position on the axis orthogonal to the direction of the goal, see \cite{M3} for details.

From the model, under the assumption that in large, low density settings, the combined effect of external influence, individual variation and ``randomness''
in the decision process may be modelled through white Gaussian noise, it is possible to derive a Langevin equation for the motion of pedestrians and a Boltzmann distribution for the pdf of pedestrian
position in a two people group. Such a distribution was used to calibrate the model by comparing with actual data collected in a pedestrian facility under conditions in agreement with the models'
assumptions.

The model, in agreement with observational data, predicts that 2 people groups:

\begin{itemize}
\item {\it are slower than individual pedestrians}, and in particular the velocity of the group is given by
\begin{equation}
\label{vg2}
v^{2} \approx v^p+\frac{2 C_\theta \eta \pi}{r_0 \kappa} < v^p,
\end{equation}
where $v^p$ is the individual preferred velocity,
\item {\it walk in an abreast formation}, the pdf for positions in centre of mass of the group following a {\it bean shaped distribution},
\end{itemize}

while $3$ people groups:

\begin{itemize}
\item are even slower, with
\begin{equation}
\label{vg3}
v^{3}\approx v^p+\frac{8 C_\theta \eta \pi}{3 r_0 \kappa} < v^{2},
\end{equation}
\item {\it walk in a V formation}, with the central pedestrian behind and the two pedestrians on the wings slightly ahead.
\end{itemize}

\section{Collision avoidance}
\subsection{Long range (circular)}
\cite{zansoc17} shows that the group model of \cite{M,M3} may be effectively combined with a collision avoidance module. More specifically, the model used in \cite{zansoc17} was derived by \cite{epl11}.
The basic idea of such model is the following: for all pedestrians we compute, with respect to every obstacle $i$ in their neighbourhood
(including moving ones, such as other pedestrians), the time $t_i$ at which they
will reach the minimum distance to such obstacle, assuming both the pedestrian and the obstacle will keep a constant velocity. We then define
\begin{equation}
t_{min}=\min_{i,\;t_i>0} t_i,
\end{equation}
and, again assuming linear motion at constant velocity,
\begin{equation}
\mathbf{r}^{min}_i=\mathbf{r}_i(t_{min})-\mathbf{r}(t_{min}),
\end{equation}
i.e., the difference between the position of the obstacle $\mathbf{r}_i$ and the of the pedestrian $\mathbf{r}$ at $t_{min}$.

Such ``future distance'' $\mathbf{r}^{min}_i$ is then used to define a the collision avoidance term of a velocity dependent specification of the social force model \cite{hel} as
\begin{equation}
  \label{fepl11}
\mathbf{F}_i=\frac{v}{t_{min}} \mathbf{F}(\mathbf{r}^{min}_i)
\end{equation}
$v$ being the magnitude of the pedestrian's velocity. While in \cite{epl11}, for historical reasons, the choice of $\mathbf{F}$ resembled the one of \cite{hel}, just replacing current
with future positions, following the analysis of \cite{thesis} in \cite{zansoc17} and in the current work we use hard core potentials
\begin{equation}
\mathbf{F}(\mathbf{r})= \begin{cases} 
A \text{ if } r \leq d_1,\\
A \frac{d_2-r}{d_2-d_1} \text{ if } d_1 < r \leq d_2,\\
0 \text{ if } r>d_2.\\  
\end{cases}
\end{equation}

\subsection{Short range (ellipse)}
We recall that according to the Social Force Model paradigm, the pedestrian decision process determines their acceleration as
\begin{equation}
  \label{sfmacc}
\ddot{\mathbf{r}}= -k_1 (\dot{\mathbf{r}}-\mathbf{v}_p) + \mathbf{F}_{int}. 
\end{equation}
Here $\mathbf{v}_p$ is the preferred velocity of the pedestrians, while $\mathbf{F}_{int}$ is the interaction (social) force\footnote{Usually when dealing with social forces masses are set to 1.}. In the model of \cite{epl11} this term is given by the sum of
over all obstacles $i$ in eq. \ref{fepl11}.

The model of \cite{epl11} was designed for moderate densities and did not take in account the shape of the human body. In order to describe
the motion of pedestrians at high density is necessary to consider at least the fact that the 2D projection of the human body is not symmetrical (this asymmetry may be first approximated by using ellipses instead of circles). When such an asymmetry is introduced, even if we still limit ourselves to the motion of pedestrians on a 2D plane, we need to introduce a new degree of freedom,
body orientation angle $\theta$. Assuming body orientation to be equal to velocity orientation would be a too strong limitation, since it would not allow pedestrians to rotate
their torso while avoiding a collision without changing considerably their motion direction. We thus consider the pedestrian to be characterised by 3 degrees of freedom, 2D position
$\mathbf{r}=(x,y)$ and angle $\theta$. The latter variable identifies the orientation of an ellipse with axes $(A, B)$ (45 and 20 cms).

As we are operating in the Social Force Model paradigm, we deal with second order differential equations for the pedestrian linear and angular acceleration $\ddot{\mathbf{r}}$ and $\ddot{\theta}$, as
functions of $\mathbf{r}$, $\dot{\mathbf{r}}$, $\theta$ and $\dot{\theta}$. Assuming the $\mathbf{r}$ and $\theta$ equations to be decoupled is highly unrealistic, since, while pedestrians are indeed
able to walk in a direction different from their body orientation, they prefer moving in the direction of their body orientation. By ignoring for the moment body oscillations due to gait, we propose
the following equations
\begin{equation}
  \label{eqmr}
\ddot{\mathbf{r}}= -k_1 (\dot{\mathbf{r}}-\mathbf{v}_p) - k_2 \Phi(\dot{\mathbf{r}},\theta) + (1-\beta(t_c)) \mathbf{F}^C + \beta(t_c) \mathbf{F}^E + \mathbf{F}^G,
\end{equation}
and
\begin{equation}
   \label{eqmt} 
\ddot{\theta}= -k_3 \dot{\theta} -k_4 \Phi(\dot{\mathbf{r}},\theta) + \beta(t_c) T^E.
\end{equation}
Many terms deserve to be explained. By $\mathbf{F}^G$ we denote the forces determined by the group potential of \cite{M3}\footnote{Whose dependence on the group member
positions is not shown for reasons of simplicity of notation.}, and by $\mathbf{F}^C$ the ``circular'' collision avoidance forces of \cite{epl11}\footnote{Again dependent of positions and velocities of
the pedestrian and of all obstacles.}. $k_2$, $k_3$ and $k_4$ are model parameter. $k_3$ accounts for the tendency of reducing body oscillations, while $k_2$ and $k_4$ for the
tendency of walking in the direction of body orientation, through the angle difference between velocity and body orientation given by $\Phi$. $t_c$ is the time at which the first collision
will happen between the ellipse representing the pedestrian body and an obstacle (such as a wall or another pedestrian), computed using an event driven algorithm \cite{eda} under the assumption
of no acceleration. The function $\beta$ introduces the idea that far away collisions are managed just using the model of \cite{epl11}, while close ones use forces taking in account body shape
($\mathbf{F}^E$, defined below). In detail, $\beta$ introduces two time scales $\tau_1$, $\tau_2$ such that 
\begin{equation}
\beta(t)= \begin{cases} 
1 \text{ if } t \leq \tau_1,\\
\frac{\tau_2-t}{\tau_2-\tau_1} \text{ if } \tau_1 < t \leq \tau_2.\\
0 \text{ if } t>\tau_2.\\  
\end{cases}
\end{equation}
As stated above, we use an event driven algorithm to compute the time of the next collision between ellipses and obstacles (polygons or other ellipses). This algorithm can be used to reproduce
the dynamics of hard ellipses undergoing elastic collisions, i.e., it provides also the force and torque that the ellipse undergoes at the moment of collision, under the assumption of
impenetrability and conservation of energy and momentum\footnote{These are actually impulsive forces and torques, i.e., expressed as a instantaneous change in linear and angular momentum.}. These
forces and momentum can be used as the basis of a collision avoidance method. Let us for simplicity consider the case of a 2D disc (i.e. ignoring $\theta$) colliding frontally with a wall
at velocity $\mathbf{v}$ in time $t$ ($t$ is the time from now at which the collision will happen if the pedestrian velocity is not modified). At the moment of collision,
the pedestrian will undergo a change in velocity $\mathbf{\Delta v}=-2\mathbf{v}$. If a force $\mathbf{\Delta v}/2t$ is applied on the pedestrian, collision will be avoided, with the pedestrian just
stopping short of the collision. By generalising to the elliptical case, we define
\begin{equation}
\mathbf{F}^E=\frac{\gamma}{t_c} \mathbf{\Delta p}
  \end{equation}
\begin{equation}
\mathbf{T}^E=\frac{\gamma}{t_c} \mathbf{\Delta L}
\end{equation}
here $\gamma$ is a model parameter, and $\mathbf{\Delta p}$ and $\mathbf{\Delta L}$ are, respectively, the linear and angular momentum exchanged at the time of collision.

Equations \ref{eqmr} and \ref{eqmt} realise the ``social'' interaction of pedestrians, i.e. their decisional process, that is, for simplicity's sake, realised at a fixed time step $\Delta t=0.05$
s (i.e., the decisional dynamics is solved by using an Euler integrator). For the physical dynamics of the system, i.e. in order to deal with {\it actual} (as opposed to predicted) collisions
between pedestrians, we use again an event driven algorithm for elastic collisions between ellipses. Such an algorithm has the positive properties to force absence of overlapping,
and thus constraints on space occupation, and to provide a quantitative measure for the amount of collisions (kinetic energy, both linear and angular, exchanged in collisions).
The negative aspect is that elastic collisions are not very realistic between pedestrians, but this problem should not be a serious one for a functional collision avoidance method, i.e.
a method in which collisions are extremely reduced.

\subsection{Optimisation}
The parameters concerning group behaviour are taken directly  from \cite{M,M3}. As stated above, the purpose of this paper is to investigate what are the possible issues of combining
a collision avoidance model with a group behaviour model, and due to the absence of data concerning both body rotation in collision avoidance and group behaviour in uncontrolled high density settings,
we decided to try first to develop an efficient collision avoidance model although not necessarily a realistic one.

We decided to use a genetic algorithm to fix the remaining parameters of the model according to the following approach. A given solution (set of model parameters) was tested on a mixed
population of individual pedestrians and pedestrians in groups (of 2 or 3 members), moving in a bi-directional flow inside a corridor with an average density of 1 pedestrian per square meter.
Using periodic boundary conditions and a fixed simulation time, for each solution we computed the following quantities\footnote{Averaged over all pedestrians and simulation time.}:
\begin{itemize}
\item Amount of energy exchanged in collisions $E$,
\item Difference between moving direction and body orientation $|\Phi|$,
\item Difference between preferred velocity and actual velocity $|\delta v|$,
\item Number of pedestrians $n$ falling at a close distance $d<0.6$ m,
\item Group potential $P$ from \cite{M} (the higher the value, the more pedestrians are far from their preferred group configuration).
\end{itemize}
It is reasonable that pedestrian may want to minimise all the above quantities, so that a possible candidate for a fitness function is
\begin{equation}
  f=-(w_E E+ w_\Phi |\Phi| + w_\delta |\delta v| + w_n n + w_P P), 
\end{equation}
Nevertheless it is not trivial to decide the values of the weights, considering also that different dimensions of the terms. The chosen solution was to define the weights in the following way.
For example, we define
\begin{equation}
w_E=\frac{1}{E_{min}}
\end{equation}
where $E_{min}$ is the optimal value of $E$ obtained by using a fitness function with the only term $f=-E$. By proceeding in a similar way for all weights, we finally obtained an adimensional fitness
function that we used to optimise our model parameters.

  \section{Experiments}
  \subsection{Experimental setting}
  In our simulations we use a corridor of width 3 meters and length 20 meters, and periodic boundary conditions. The width of the corridor is chosen is such a way that two 3 people groups walking in opposite directions may not
  cross without changing their formation, based on the group spatial size as reported in \cite{M}. We use four different density conditions, $\rho=1$ ped/m$^2$, i.e. 60 pedestrians,
  $\rho=2$ ped/m$^2$, i.e. (120 pedestrians), $\rho=3$ ped/m$^2$, (180 pedestrians) and $\rho=4$ ped/m$^2$ (240 pedestrians). We also use two different group rate $r_g$ conditions, namely without groups $r_g=0$ and
  with half of the pedestrians in groups, $r_g=0.5$. When groups are present, 20\% of them are in 3 people groups, and 30\% of them in 2 people groups.
  We thus have 8 conditions depending on the values of $\rho$ and $r_g$. For example, when $\rho=4$ ped/m$^2$ and $r_g=0.5$ we have 120
  individual pedestrians, 36 2 people groups and 16 3 people groups. Initial conditions are determined by dividing the corridor in cells of equal size, and placing pedestrians in
  randomly chosen cells, regardless of their flow direction\footnote{Pedestrians in the same group are nevertheless located in neighbouring cells.}. The pedestrians' preferred velocities are chosen from
  a normal distribution with $\mu=1.2$m/s and $\sigma=0.2$ m/s \footnote{Pedestrians in the same group have the same preferred velocity.}. Each group or pedestrian has a probability 0.5 of belonging to each one of the flows\footnote{Thus flows have the same weight only in average.}. For each condition, 10 different simulations with independent initial conditions are used and the observables
  defined below are averaged over these initial conditions.
  \subsection{Observables}
  For each experimental condition, we define the following observables:
  \begin{itemize}
  \item rate of velocity over preferred velocity,
    \begin{equation}
\nu=\frac{v}{v_p}
    \end{equation}
  \item energy exchanged in collisions $E$,
  \item number of lanes $n_l$,
  \item rate of pedestrians in lanes
    \begin{equation}
r_l=\frac{\sum_{i \in n_l} s_i}{N},
    \end{equation}
    where $s_i$ is the number of pedestrians in lane $i$ and $N$ is the overall number of pedestrians.
  \end{itemize}
  Lane recognition is performed using the algorithm described below.

  For each observable we first compute the average over pedestrians and time for each independent initial condition, and then compute the average, standard deviation and standard error over the different independent conditions. The latter are the values shown in the figures. In order to show also time dependence, time averages are computed over 10 slots of length $T_i=20$ s.

\subsection{Lane clustering algorithm}
For the analysis of the lane formation phenomenon we have applied the algorithm discussed in~\cite{Crociani2018AIIA}. As a novel part of this paper, we have introduced an extended version of the algorithm which is able to produce more stable results in dense and possibly chaotic situations. 

\begin{figure}[t]
\begin{center}
\includegraphics[width=.99\textwidth]{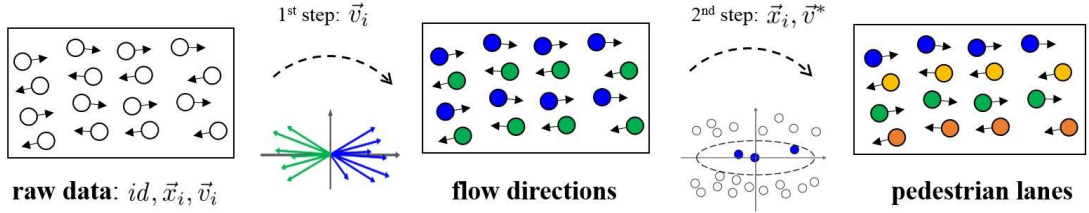}
\caption{Workflow of the hierarchical clustering algorithm to characterize the lane formation.}
\label{fig:clustering_workflow}
\end{center}
\end{figure}

The algorithm is based on a hierarchical two-steps application of DBSCAN~\cite{DBSCAN}, with distance metrics and respective parameters specifically tailored to deal with this problem. The aim is to achieve clusters in tune with the intuitive conception of the lane formation phenomenon. The workflow is described in Fig.~\ref{fig:clustering_workflow}.

The choice of a hierarchical approach is mainly motivated by the need of knowing the average flow direction before the final identification of lanes: this information is in fact used to identify clusters which describe queuing pedestrians in the observed scenario. Moreover, we assume that the parameter $minPoints$ --describing the smallest size of a cluster in DBSCAN-- is set to 3 in both steps, since we define three persons walking in a river-like pattern as the simplest case of lane. The algorithm works on almost instantaneous data, potentially allowing the implementation on real-time systems: with the usage of a de-noising algorithm, smoothed positions and velocity vectors of pedestrians related to small time-windows ($\textless$ 0.5s) can be calculated (in this work we did not use particular smoothing methods since the outcome positions from the simulation model are already sufficiently clean).

The two steps of the algorithm are defined to compute clusters describing respectively: \textit{(i)} the dominant flow directions in the observed scenario; \textit{(ii)} the actual lanes of pedestrians.
The dominant flow directions are achieved by aggregating velocity vectors $\vec{\nu_i}$ of pedestrians, using the cosin dissimilarity as distance metric. This metric describes the angle $\alpha_{ij}$ between two velocity vectors: 

\begin{equation}
\alpha_{ij} = arccos\left(\frac{\vec{v_i}\cdot\vec{v_j}}{\|\vec{v_i}\| \cdot \|\vec{v_j}\|} \right)
\end{equation}

Two vectors $\vec{\nu_i}$ and $\vec{\nu_j}$ are considered \textit{neighbours} if $\alpha_{ij} \leq \Theta_\nu$. Following the logic of DBSCAN, a point is then marked as \textit{core} if it has at least $\mathit{minPoints}$ neighbours, as \textit{border} if it is neighbour of a core point or \textit{noise} otherwise. A cluster is finally the set of all neighbour core and border points.

The output clusters of this first step are \textit{dense} set of vectors, which should be representative of a direction of motion. This is verified in most of the cases, but it can happen that in particularly chaotic scenarios, where the observed crowd of pedestrian shows many patterns of movement directions, the whole set of vectors gets aggregated in a unique cluster. The output in this case is not representative of any direction of movement. 

The problem can be overcome by acting on the parameters $\Theta_\nu$ and $\mathit{minPoints}$. These two define the value of density searched in the dataset, thus it is possible to find a combination of parameters that leads to a proper segmentation of vectors among representative clusters. Acting statically on parameters is, though, a double-sided weapon: one might find a configuration which works perfectly with very dense and chaotic scenarios, but lacks proper results in more simple situations. 

With these considerations, we configured an extension that iteratively adjusts the parameter $\mathit{minPoints}$ for the current step analysed with the algorithm\footnote{Note that at the beginning of each frame analysed with the algorithm, $\mathit{minPoints}$ is reset to its initial value of 3.}, to further split clusters that do not satisfy a correctness criterion. The new version is described in Alg.~\ref{alg:iterative_search}, assuming $\mathit{Points}$ to be the set of all velocities at the beginning of each frame and $\mathit{max\_it}$ being the maximum number of iterations allowed to refine the result.

\begin{algorithm}[t]
\caption{Iterative refining of velocity clusters.}\label{alg:iterative_search}
\begin{algorithmic}[1]
\While {$\mbox{Points} \neq \emptyset \mathbf{\ and\ } \mbox{max\_it} > 0$}
	\State $\mbox{Vel\_clusters} \leftarrow \mathit{compute\_vel\_clusters}(\mbox{Points},\Theta_\nu,\mbox{minPoints})$
	\State $\mbox{Points} \leftarrow \emptyset$ 
	\For {$\mbox{Cluster} \in \mbox{Vel\_clusters}$}
		\If {$\mathit{is\_representative}(\mbox{Cluster})$}
			\State $\mbox{Results} \leftarrow \mbox{Results} \cup \left\lbrace\mbox{Cluster}\right\rbrace$
		\Else
			\State $\mbox{Points} \leftarrow \mbox{Points} \cup \mbox{Cluster}$
		\EndIf
	\EndFor
	\State $\mbox{minPoints} \leftarrow \mbox{minPoints} + \Delta_{\mathit{points}}$	
	\State $\mbox{max\_it} \leftarrow \mbox{max\_it} - 1$
\EndWhile
\State \Return $\mbox{Results}$
\end{algorithmic}
\end{algorithm}

In this way, for a maximum of $\mathit{max\_it}$ iterations, the algorithm will search for denser clusters in the dataset until all achieved clusters do satisfy the $\mathit{is\_representative}$ criterion. The increase of density among iterations is regulated via the $\Delta_{\mathit{points}}$ parameter. Moreover, points in clusters that do not satisfy the criterion after $\mathit{max\_it}$ iterations will be marked as noise. As for the correctness criterion, we configured a threshold on the standard deviation of velocities inside each cluster, so that to define a maximum width of its distribution of velocities.

For the analysis later presented, we calibrated manually the algorithm and we set the parameters as $\left( \mathit{minPoints}, \Theta_\nu, \mathit{max\_it}, \Delta_{\mathit{points}}\right) = \left(3, 10^\circ, 5, 2\right)$. Results highlighted a significant improvement achieved with this extension.

The second step reflects the definition provided in~\cite{Crociani2018AIIA}. It works sequentially on individual clusters previously identified, by using a distance metric able to characterize \emph{queueing} pedestrians. This step analyses pedestrians positions and the average velocity $\vec{V}_{C^*}$ (Eq.~\ref{eq:avg_velocity}) of the associated cluster $C^*$ found with the previous step, if any. The applied distance metric $\Lambda_{ij}$ aims, in fact, at aggregating points following the direction of movement. With this purpose, the metric defines an elliptical neighbourhood, in which the ellipse is rotated in order to have its long side parallel to the direction of motion described by $\vec{V}_{C^*}$. This is formalized in the following equations:

\begin{equation}
\Lambda_{ij} = \sqrt{\left( \cfrac{\hat{x}_{ij}}{\Xi_x} \right)^2 + \left(\hat{y}_{ij}\right)^2}
\end{equation}

\begin{equation}
(\hat{x}_{ij},\hat{y}_{ij}) = \bigodot\left(\vec{x_j} - \vec{x_i}, -\measuredangle\left(\vec{V}_{C^*},\left(1,0\right)\right)\right)
\end{equation}

\begin{equation}
\measuredangle(\vec{\nu_i}, \vec{\nu_j}) = 
\begin{cases}
\alpha_{ij} & \mbox{if } \mathit{DET}(\vec{\nu_i}, \vec{\nu_j}) \leq 0\\
360 - \alpha_{ij} & \mbox{otherwise}
\end{cases}
\end{equation}

\begin{equation}\label{eq:avg_velocity}
\vec{V}_{C^*} = \displaystyle\sum_{j \in C^*} \left( \frac{\vec{\nu_j}}{|C^*|} \right) 
\end{equation}

Here $\bigodot$ rotates counter-clockwise the vector $\vec{x_j}$ w.r.t $\vec{x_i}$ by the degrees described with the second input. $\measuredangle$ computes the clockwise angle between the two input vectors ($\mathit{DET}$ is the determinant of the matrix built with $\vec{\nu_i}$ and $\vec{\nu_j}$). Regarding this passage, we configured the parameters as $\left(\mathit{minPoints}, \Theta_x, \Xi_x\right) = \left(3, 0.8m, 3\right)$.

\section{Results}
\begin{figure}[ht!]
\begin{center}
\includegraphics[width=0.7\linewidth]{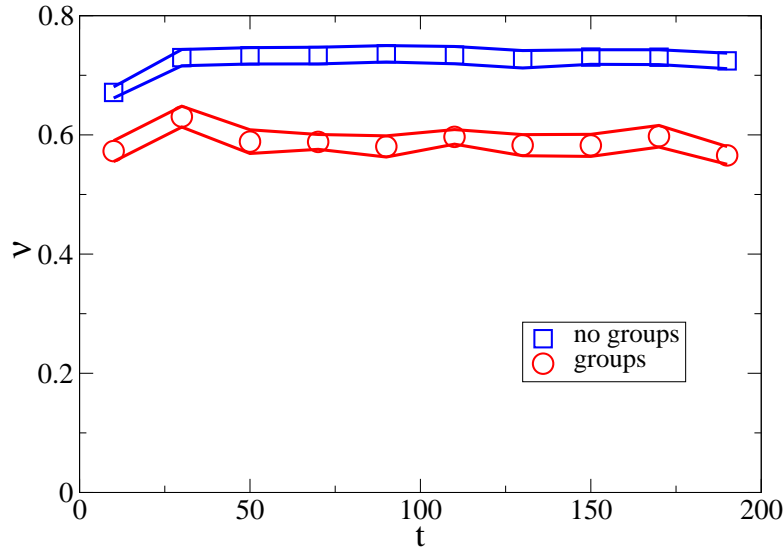} 
\caption{$\nu(t)$ for the $\rho=1$ ped/m$^2$ condition. Blue and squares correspond to $r_g=0$, while red and circles to $r_g=0.5$. Continuous lines correspond to standard error confidence intervals.}
\label{f1}
\end{center}
\end{figure}

\begin{figure}[ht!]
\begin{center}
\includegraphics[width=0.7\linewidth]{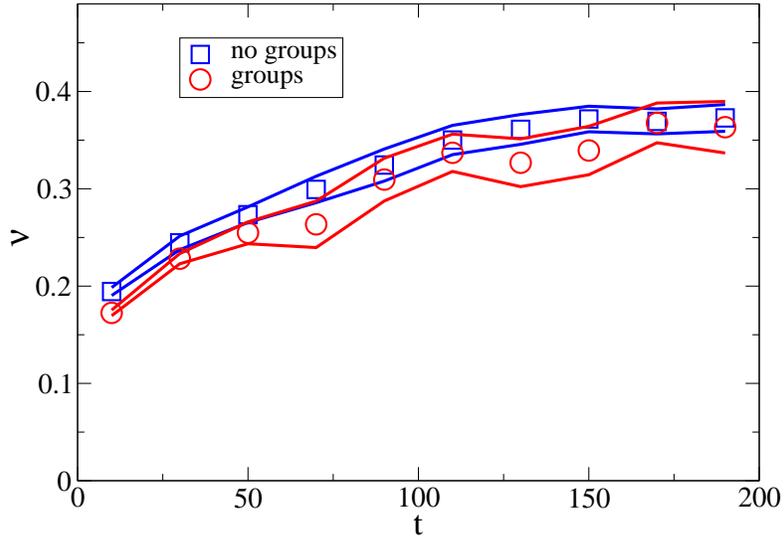} 
\caption{$\nu(t)$ for the $\rho=4$ ped/m$^2$ condition. Blue and squares correspond to $r_g=0$, while red and circles to $r_g=0.5$. Continuous lines correspond to standard error confidence intervals.}
\label{f2}
\end{center}
\end{figure}

Figure \ref{f1} shows the time evolution of the $\nu$ observable in the $\rho=1$ ped/m$^2$ condition, while Figure \ref{f2} shows the corresponding result in the $\rho=4$ ped/m$^2$ condition. We may observe that the effect of groups on velocity is very strong in the medium
density range, in which flows without groups present a clearly higher velocity. The velocity difference is higher than the one between groups and individuals in moderate density settings \cite{M2}. On the other hand, the effect of the presence of groups
on velocity is not present at very high densities.

\begin{figure}[ht!]
\begin{center}
\includegraphics[width=0.75\linewidth]{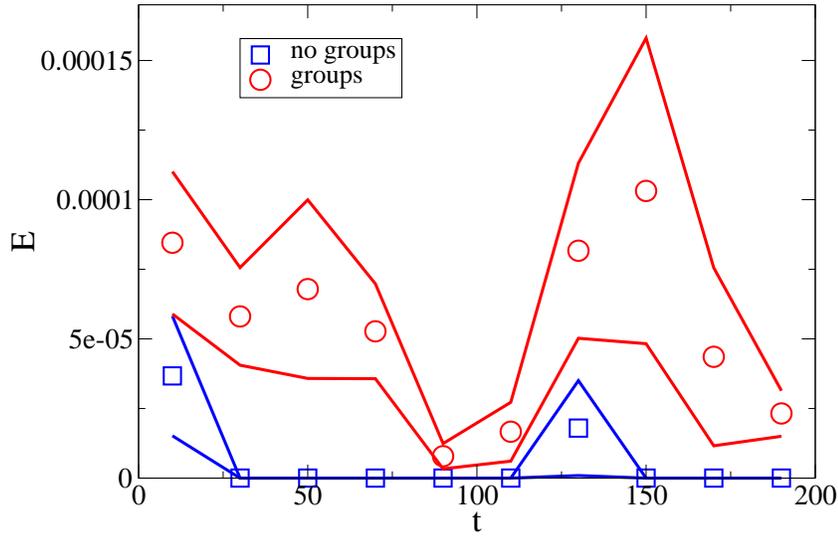} \hspace{0.07\linewidth}
\caption{$E(t)$ for the $\rho=1$ ped/m$^2$ condition. Blue and squares correspond to $r_g=0$, while red and circles to $r_g=0.5$. Continuous lines correspond to standard error confidence intervals. $E$ is measured as kinetic energy corresponding to a mass 1 exchanged in average by a pedestrian each second.}
\label{f3}
\end{center}
\end{figure}

\begin{figure}[ht!]
\begin{center}
\includegraphics[width=0.7\linewidth]{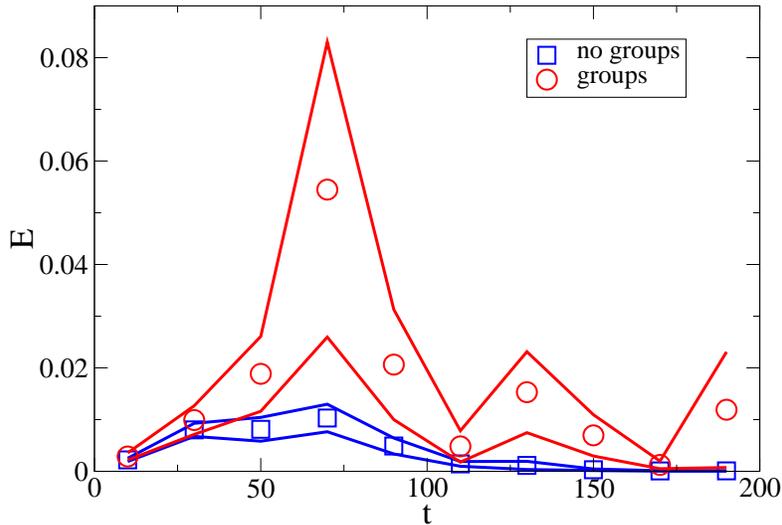} 
\caption{$E(t)$ for the $\rho=4$ ped/m$^2$ condition. Blue and squares correspond to $r_g=0$, while red and circles to $r_g=0.5$. Continuous lines correspond to standard error confidence intervals. $E$ is measured as kinetic energy corresponding to a mass 1 exchanged in average by a pedestrian each second.}
\label{f4}
\end{center}
\end{figure}

Figure \ref{f3} shows the time evolution of the $E$ observable in the $\rho=1$ ped/m$^2$ condition, while Figure \ref{f4} shows the corresponding result in the $\rho=4$ ped/m$^2$ condition. We may observe that the amount of collision is strongly increased in presence of groups for any value of $\rho$.

\begin{figure}[ht!]
\begin{center}
\includegraphics[width=0.7\linewidth]{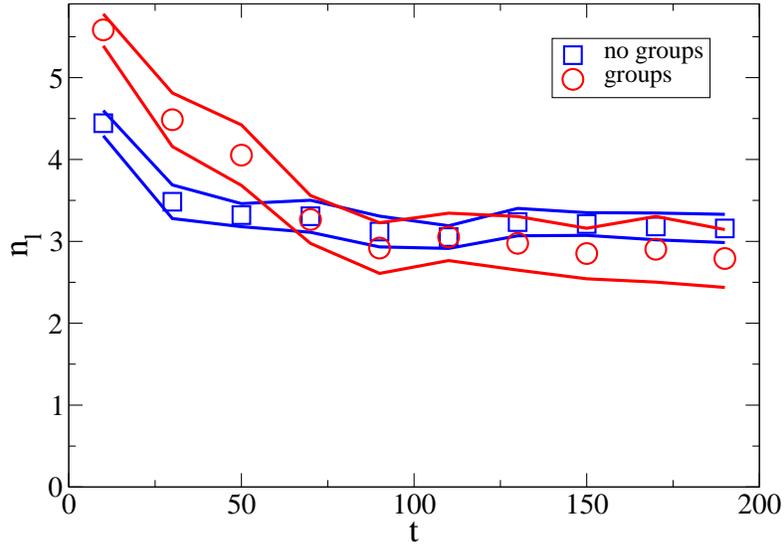} 
\caption{$n_l(t)$ for the $\rho=2$ ped/m$^2$ condition. Blue and squares correspond to $r_g=0$, while red and circles to $r_g=0.5$. Continuous lines correspond to standard error confidence intervals.}
\label{f5}
\end{center}
\end{figure}

\begin{figure}[ht!]
\begin{center}
\includegraphics[width=0.7\linewidth]{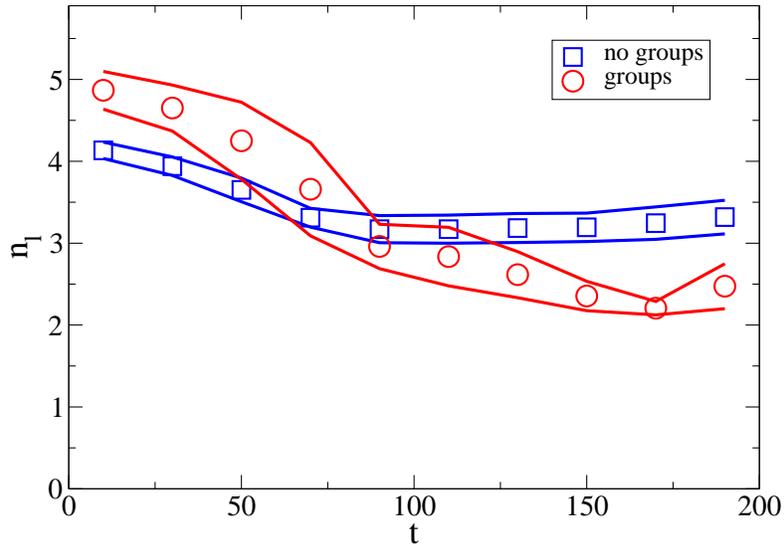} 
\caption{$n_l(t)$ for the $\rho=3$ ped/m$^2$ condition. Blue and squares correspond to $r_g=0$, while red and circles to $r_g=0.5$. Continuous lines correspond to standard error confidence intervals.}
\label{f6}
\end{center}
\end{figure}

\begin{figure}[ht!]
\begin{center}
  \includegraphics[width=0.7\linewidth]{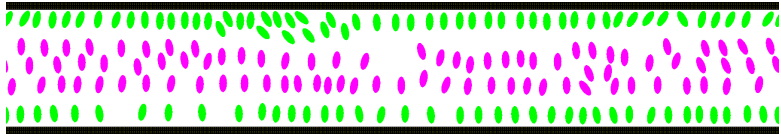}
\caption{Simulation snapshot for $\rho=3$ ped/m$^2$ and $r_g=0$. Green move to the left; violet to the right.}
\label{f7}
\end{center}
\end{figure}

\begin{figure}[ht!]
\begin{center}
   \includegraphics[width=0.7\linewidth]{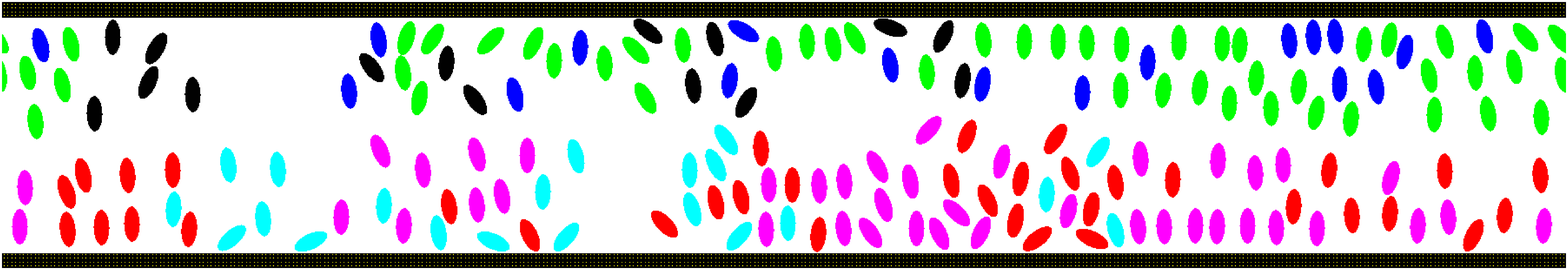} 
\caption{Simulation snapshot for $\rho=3$ ped/m$^2$ and $r_g=1$. Green move to the left; red, violet (dyads) and cyan (triads) to the right.Green, black (triads) and blue (dyads) move to the left; violet, red (dyads) and cyan (triads) to the right.}
\label{f8}
\end{center}
\end{figure}

Figure \ref{f5} shows the number of lanes in the $\rho=2$ ped/m$^2$ condition in absence and presence of groups. Although the number of lanes stabilises around 3 in both cases, in the initial stage
the flows with groups present an higher number of lanes. The corresponding result in the $\rho=3$ ped/m$^2$ condition is shown in Figure \ref{f6}. In this case also the final number of lanes is different, converging to 3 in absence of groups,
and 2 in presence of groups (see also figures \ref{f7}, \ref{f8}; the presence of groups occupying a wider space makes the formation of three lanes more difficult).


\begin{figure}[ht!]
\begin{center}
\includegraphics[width=0.7\linewidth]{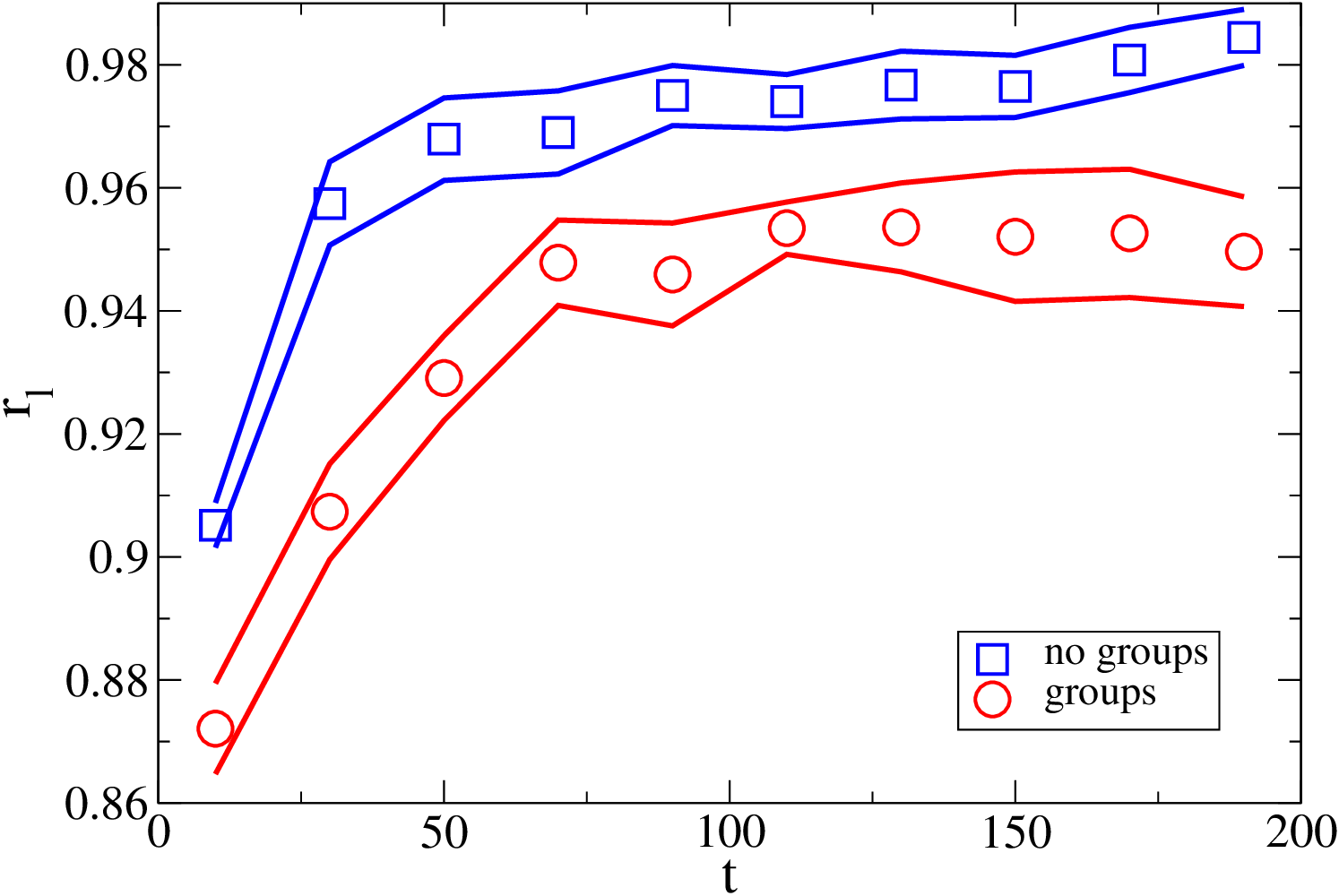} 
\caption{$r_l(t)$ for the $\rho=2$ ped/m$^2$ condition. Blue and squares correspond to $r_g=0$, while red and circles to $r_g=0.5$. Continuous lines correspond to standard error confidence intervals.}
\label{f9}
\end{center}
\end{figure}

Finally, Figure \ref{f9} shows the rate of pedestrians in lanes in the $\rho=2$ condition in absence and presence of groups. As it happens in general, a larger number of pedestrians is organised in lanes
when groups are absent.

  \section{Conclusions}
  We have verified that by combining a realistic model for group behaviour at moderate crowd density with an efficient collision avoidance model, the presence of groups has
  a strong impact on the dynamics of bi-directional flows, affecting the velocity of pedestrians, the amount of collisions between pedestrians, the number of lanes and the rate of pedestrians
  organised in lanes.

  By trying to move on specific formations to enhance not only proximity but also communication between the group members (abreast and V formations) groups occupy a larger portion of the corridor
  and have less moving flexibility. Their presence decreases the average velocity of the crowd, makes organisation in lanes more difficult and increases the number of collisions.

  We have no claim that these results reflect the reality of the effect of groups on crowd dynamics. Although the impact on velocity and lane organisation goes in the expected direction and seems
  reasonable at least from a qualitative point of view,
  it is questionable that crowds with groups present such a larger amount of collisions. Just combining a collision avoidance and a group behaviour model could overlook important aspects such as
  a specific behaviour of pedestrians {\it towards} groups, and a density-dependent tendency of groups to give up communication and thus spatial formations to avoid collisions.

  We nevertheless believe that our preliminary results seriously hint at problems and difficulty of a naive approach to the presence of groups in crowds.

\section{Acknowledgements}
  
  This research is partially based on results obtained from a project
commissioned by the New Energy and Industrial Technology
Development Organization (NEDO).

\end{document}